# An improved 2.5 GHz electron pump: single-electron transport through shallow-etched point contacts driven by surface acoustic waves

P. Utko, K. Gloos, J. B. Hansen, and P. E. Lindelof

Nano-Science Center, Niels Bohr Institute fAFG, University of Copenhagen, Universitetsparken 5, DK-2100 Copenhagen, Denmark

We present an experimental study of a 2.5 GHz electron pump based on the quantized acoustoelectric current driven by surface acoustic waves (SAWs) through a shallow-etched point contact in a GaAs/AlGaAs heterostructure. At low temperatures and with an additional counter-propagating SAW beam, up to $n = 20$ current plateaus at $I = nef$ could be resolved, where $n$ is an integer, $e$ the electron charge, and $f$ the SAW frequency. In the best case the accuracy of the first plateau at 0.40 nA was estimated to be $\Delta I/I = \pm$ 25 ppm over 0.25 mV in gate voltage, which is better than previous results.

PACS numbers: 73.63.Kv, 72.50.+b, 06.20.Jr

# 1. Introduction

Recently a new approach towards single-electron manipulation has been demonstrated, where surface acoustic waves (SAWs) are used to transfer electrons across the potential hill of a quantum point contact (QPC) defined in a GaAs/AlGaAs heterostructure [1-5]. At gate voltages below pinch-off for the conductance, when the contact is still closed, the acoustoelectric current develops plateaus at $I = nef$. Here $e$ is the electron charge, $f$ the SAW frequency, and $n$ an integer. Qualitatively, this situation can be described by "moving quantum dots" travelling with the sound velocity [1-2]: Electrons are trapped in local minima of the dynamic SAW potential and transferred across the potential barrier of the point contact. Coulomb repulsion between electrons in such travelling minima determines the occupancy of the dots within some interval of gate voltage, SAW magnitude, and SAW frequency.

The typical driving frequency of a SAW-based device (~2.5 GHz) is significantly higher than the maximum operation frequency of other existing electron pumps [6-8]. The latter devices operate at frequencies up to several 10 MHz, resulting in an output current of less than 10 pA. This limits their prospective applications as a quantum standard of electric current. On the contrary, a high-frequency SAW pump delivers quantized current in the 1 nA range – sufficiently high for metrological applications. Its accuracy, however, is limited by the finite slope of the acoustoelectric current plateaus as function of gate voltage. This makes it difficult to find the exact value of the quantized current. Possible mechanisms responsible for such a deviation from the ideal value of $I = nef$, mainly electron co-tunneling and thermal activation, have been discussed in a number of theoretical [9-11] and experimental studies [1-5]. The precision of the quantization was significantly improved when the point contact was patterned by the shallow-etch technique [3], or when a second counterpropagating SAW beam was used in devices with split-gate defined constrictions [4].

We have combined these two techniques to improve the flatness of the plateaus, by investigating the quantized acoustoelectric current through shallow-etched point contacts with two counterpropagating SAW beams. This allowed us to vary the relative amplitude and phase of both SAW beams, as well as to take advantage of the properties of the shallow-etched constrictions: their well-defined shape and large subband separation [12]. In this way we could observe up to 20 plateaus in the acoustoelectric current at low temperatures in high-mobility two-dimensional electron gas (2DEG) in GaAs/AlGaAs heterostructures.

## 2. Experiment

Figure 1a shows the sample layout. Two interdigital transducers (IDTs), facing each other, generate the surface acoustic waves. The IDTs were deposited 2.6 mm apart on both sides of a 2DEG mesa with a quantum point contact in the center. The interdigital electrode spacing set the fundamental acoustic wavelength of the transducers to about 1.2 μm, and thus their center frequency to around 2.45 GHz. The GaAs/AlGaAs heterostructure had a mobility of 105 $m^2/Vs$ and a carrier density of $2.8 \times 10^{15}$ $m^{-2}$, measured in the dark at 10 K. The QPC was patterned by e-beam lithography. Two semicircular shallow-etched trenches formed a smooth constriction between the two electron reservoirs, whereas the large areas of the 2DEG across the channel served as side gates (Fig. 1b). The trenches had a curvature radius of 10 μm. They were 200 nm wide and 40 nm deep.

Our experimental setup was similar to that used by Cunningham *et al.* [4]. The rf signal from the microwave generator was split up and simultaneously applied to both transducers. Due to the phase shifter and the attenuator in one of the rf lines (Fig. 1c), the relative magnitude and the relative phase of both signals could be varied and adjusted. A low-noise current preamplifier detected the acoustoelectric current. All measurements were carried out at 1.2 K. Applying the microwave excitation to the transducers increased the temperature reading up to about 1.7 K.

For sufficiently high rf-power levels, the acoustoelectric current can be observed below pinch-off voltage for conductance. Figure 2 shows how this current changes with gate voltage. The slope of the current plateaus depends on gate voltage, frequency, the rf power as well as on the relative magnitude and phase of the microwave signals at the two transducers. To help optimizing these parameters for a minimum slope, a small 117 Hz modulation of $dV_g \leq 1$ $mV_{rms}$ was added to the gate voltage and the ac component of the acoustoelectric current, $dI$, was monitored with a lock-in amplifier. Figure 2 also shows the directly measured slope $dI/dV_g$. After optimization, the slope of the first plateau was reduced by nearly an order of magnitude with respect to the case when the second beam was off. According to Figure 2, up to 20 plateaus appeared in $I(V_g)$, although only about half of them are well-pronounced while the other half are rather weak. The obviously clear distinction between the two types of plateaus could be due to the type of spatial ordering of electrons in the moving quantum dots. Figure 3 shows the position of the current plateaus with respect to the first plateau for the different power levels on the second IDT. The number of plateaus per gate voltage increases strongly when the power is increased.

## 3. Discussion

The two counterpropagating SAW beams can be described as a superposition of a travelling wave and a (usually weaker) standing wave. The position of its nodes and its magnitude can be varied with respect to the constriction by changing the phase (or the frequency). The dynamical tuning of the traveling SAW potential minima by the standing wave can be particularly important for electron transport when a node sits at the rising slope of the QPC potential hill or near an impurity which could be at the entrance to the constriction.

Figure 4 shows the most precise quantization of the first plateau obtained by us so far. 200 readings of the current were taken over 5 seconds at constant gate voltage. Then the gate voltage was increased by 2.5 μV and another data set taken. For each of the data points obtained in this way, the standard deviation of the mean current was calculated and plotted as error bars. The location of the quantized current was found by a two-step procedure. First, the derivative of $I(V_g)$ was calculated and a ~1 mV wide region around the minimum slope was chosen for further inspection. On average, this minimum slope was quite small, only 0.06 pA/mV. Second, the flattest part over this range was selected. This turned out to be 0.25 mV wide, containing about 100 data points. If we attribute this flat region to the ideal quantized current $I = ef$, we get a relative accuracy of $\Delta I/I = \pm 25$ ppm. The absolute value of the experimental quantized current of 394.695 pA differed by 0.21% from the theoretical value of $ef$ = 393.855 pA. This was within the accuracy of the current amplifier and AD converter of about ± 1 pA.

## 4. Summary

To conclude, we have investigated the single-electron transport through shallow-etched point contacts driven by two counterpropagating SAW beams. Through proper optimization of the electrical parameters up to 20 plateaus appeared in the acoustoelectric current. The highest accuracy of the first plateau was estimated to be $\pm 25$ ppm over a gate voltage interval of 0.25 mV, which is a considerable improvement compared to the previous results.

We acknowledge support from European Commission FET Project SAWPHOTON.

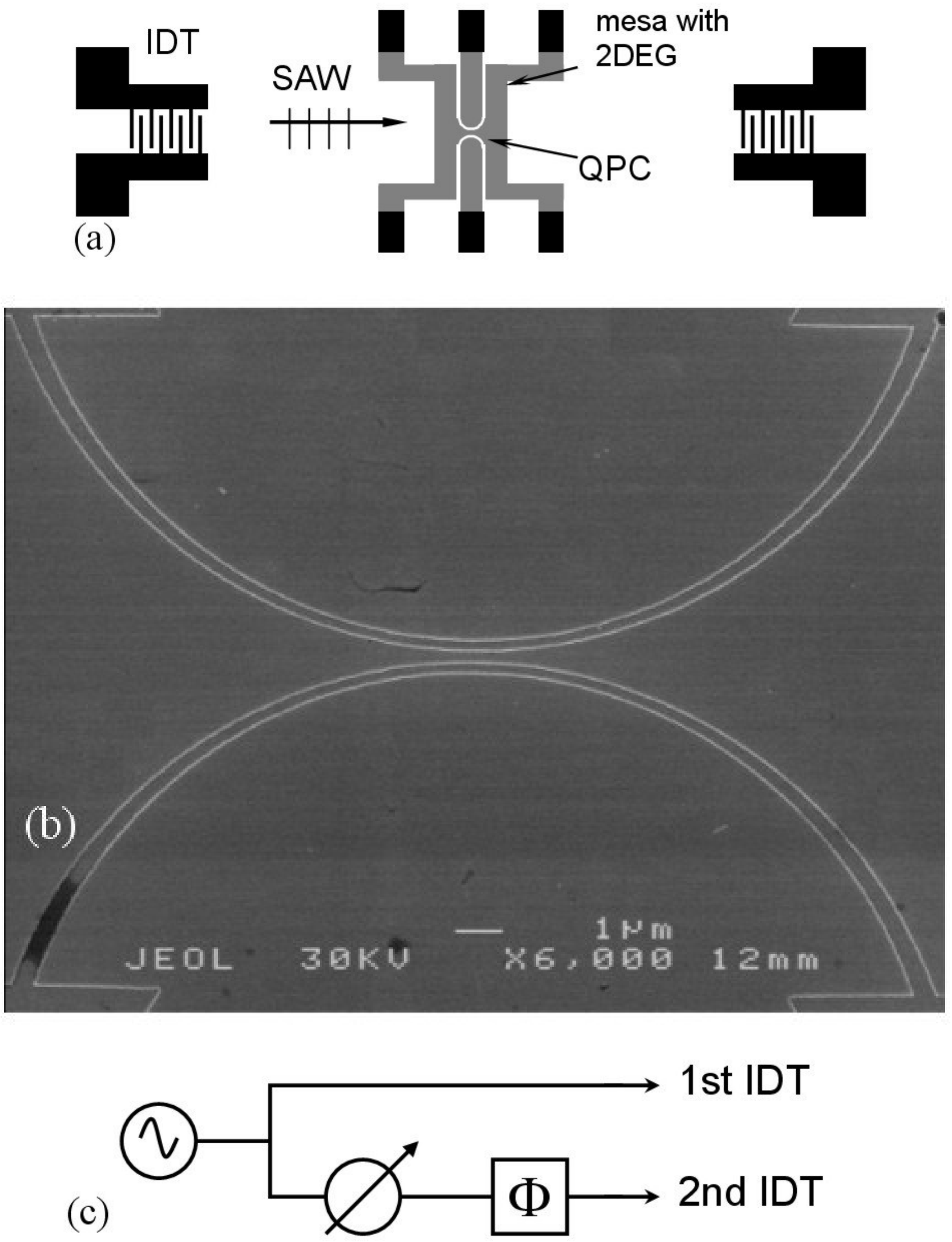

Fig. 1.

Experimental setup. (a) Sample layout with 2DEG mesa, quantum-point contact (QPC), side gates, and interdigital transducers (IDTs). (b) SEM picture of the QPC. (c) Schematic of the rf system with generator, beam splitter, attenuator, and phase shifter in the rf line to the 2nd IDT.

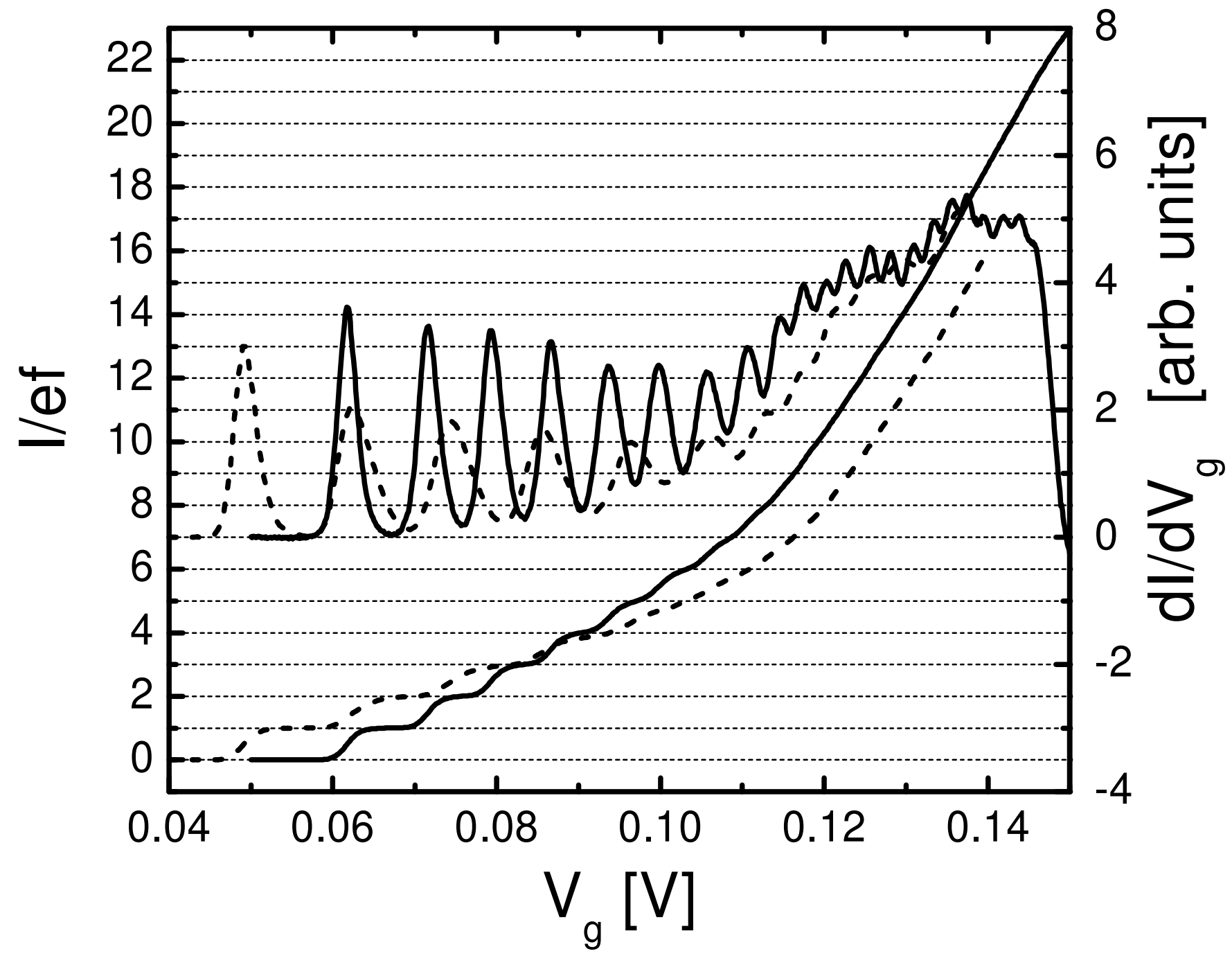

Fig. 2.

Acoustoelectric current $I$ normalized to $ef$ = 394.796 pA and its derivative $dI/dV_g$ versus gate voltage $V_g$ at T = 1.2 K. The rf power of 16.5 dBm at f = 2464.125 MHz was applied to the port of the first IDT. The excitation applied to the port of the second IDT was reduced by 8.5 dB (solid line) and 20.5 dB (dashed line), respectively. The conductance pinch-off was at 0.145 V when the rf was off.

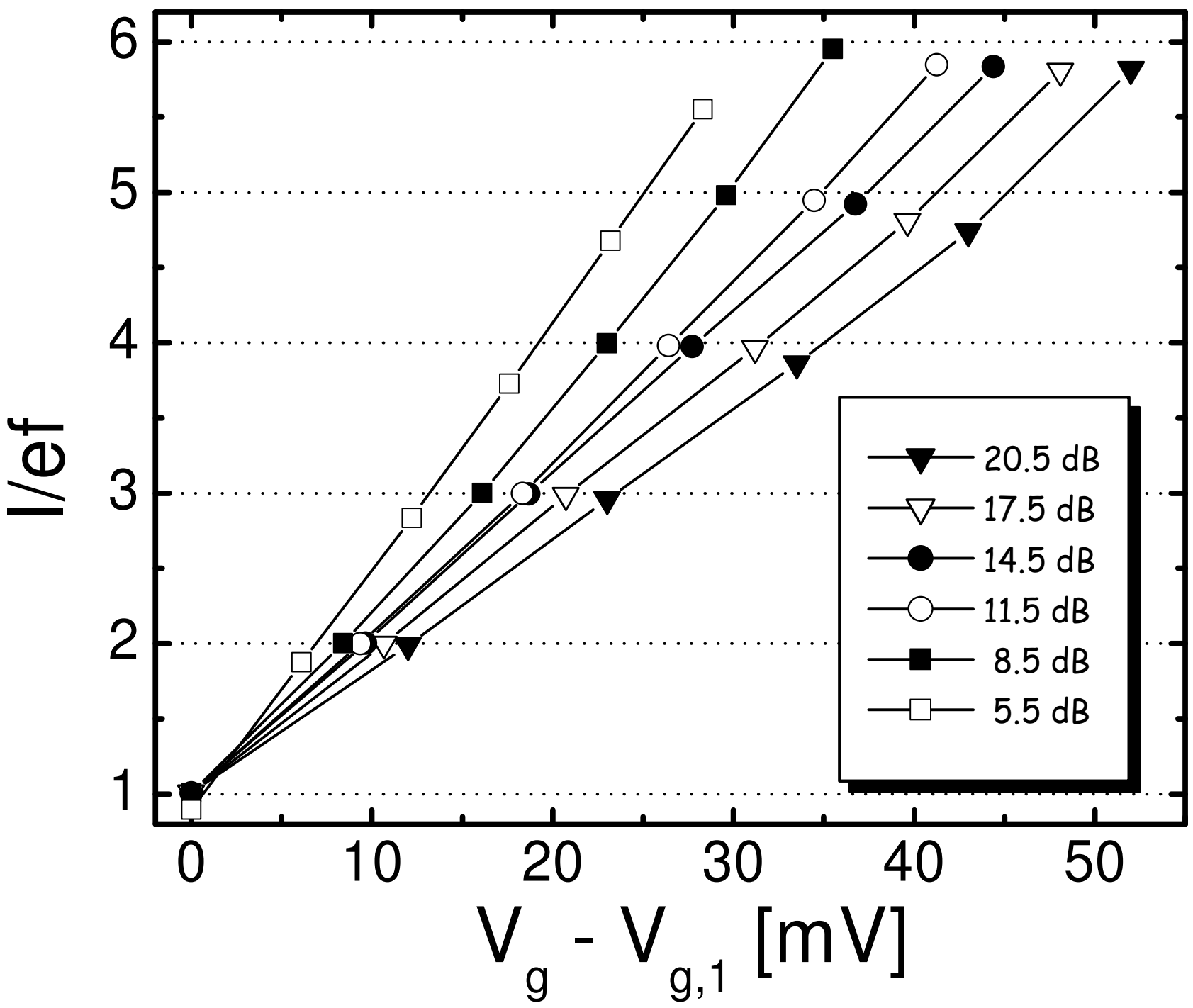

Fig. 3.

Position of the acoustoelectric current plateaus at minimum slope $dI/dV_g$ for the indicated attenuation of the counter-propagating beam. The rf power on the port to the first transducer was 16.5 dBm at f = 2464.125 MHz. The position of the first plateau at gate voltage $V_{g,1}$ depends on the applied power.

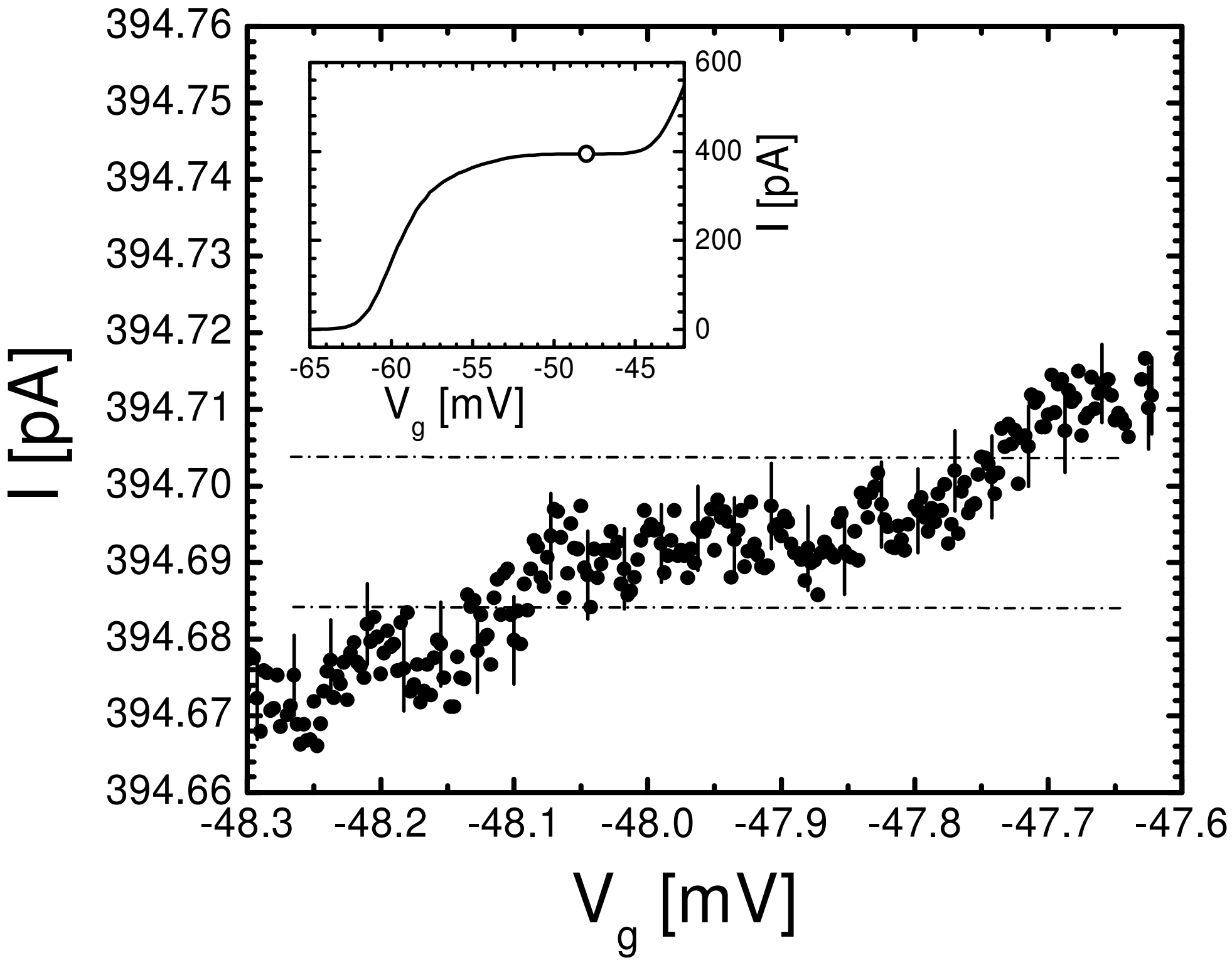

Fig. 4.

Acoustoelectric current $I$ versus gate voltage $V_g$ at the optimum settings. The SAW frequency was $f = 2458.250$ MHz. Only the data points around the minimum slope are shown. Error bars represent the standard deviation of the current reading. The two horizontal dashed lines comprise a $\Delta I/I = \pm 25$ ppm interval. Inset: Full $I(V_g)$ curve with a mark at the flattest part.